# In-plane and out-of-plane electric dipoles and phase transitions in 2D-layered TlGaS$_2$

A. D. Molchanova[1], L. H. Yin[2,*], L. P. Gao[2], W. H. Song[2], Y. P. Sun[2,3], K. R. Allahverdiyev[4], and M. N. Popova[1,§]

[1] *Institute of Spectroscopy, Russian Academy of Sciences, Troitsk, Moscow 108840, Russia*

[2] *Key Laboratory of Materials Physics, Institute of Solid State Physics, HFIPS, Chinese Academy of Sciences, Hefei 230031, People's Republic of China*

[3] *High Magnetic Field Laboratory, Chinese Academy of Sciences, Hefei 230031, People's Republic of China*

[4] *National Aviation Academy, Baku AZ1045, Azerbaijan*

**Abstract**

Out-of-plane and in-plane electric polarization, which rarely coexist in a two-dimensional (2D) ferroelectric material, offer different advantages in ferroelectricity-based devices. Here, we report the coexistence of in-plane and out-of-plane electric dipoles, along with various phase transitions, in 2D van der Waals layered TlGaS$_2$ single crystal. Quantum paraelectricity was observed along both in-plane and out-of-plane directions of the TlGaS$_2$ crystal. Detailed investigation of the quantum paraelectric soft-mode behavior reveals a close correlation between the electric dipoles and the off-center displacement of Tl$^{1+}$ ions with 6s$^2$ lone pairs in TlGaS$_2$. Anomalies near temperatures of about 120 K and 60-75 K in dielectric and/or infrared spectra indicate the existence of local or weak long-range structural transitions in TlGaS$_2$. Our results provide important experimental evidence for elucidating the phase transitions and coexistence of in-plane and out-of-plane electric dipoles in 2D layered TlGaS$_2$.

\* Corresponding author: E-mail: lhyin@issp.ac.cn

§E-mail: popova@isan.troitsk.ru

# I. INTRODUCTION

Ferroelectric (FE) materials, characterized by spontaneously polarized electric dipoles that can be switched by an applied electric field (E), have attracted sustained attention due to their potential applications in nonvolatile information storage, transducers, nonlinear optical devices, and beyond [1,2]. In contrast to ferroelectricity in conventional bulk materials, FE orders in 2D van der Waals layered materials exhibit fundamentally new physics and phenomena [3,4]. Ferroelectricity at the atomic thickness limit in 2D FE materials breaks traditional downscaling barriers and enables novel polarization switching mechanisms [4]. For example, 2D $Bi_2O_2Se$ films have been shown to exhibit robust out-of-plane ferroelectricity at room temperature (RT) even in the single-unit-cell thickness limit, with FE polarization proposed to arise from electric-field-modulated dipole ordering associated with Se atom displacements [5]. In fact, out-of-plane polarization is the dominant configuration in modern commercial FE devices such as FE random-access memory (FeRAM), owing to its advantages in high storage density and low operating voltage. In contrast, 2D layered FE materials like monolayer SnS have recently been found to exhibit interesting in-plane polarization [3], which offers excellent complementary metal-oxide semiconductor (CMOS) compatibility and simpler planer electrode fabrication and is favorable for FE field-effect transistors (FeFET) based devices [6]. Departing from the Landau–Ginzburg double-well potential in classical FEs, an abnormal quadruple-well potential and associated mobile Cu atoms have been reported in 2D layered FE $CuInP_2S_6$ (CIPS) crystals [7]. Coexistence of ferroelectricity and metallicity have been observed in 2D $WTe_2$ [8]. These emerging functionalities make FE 2D materials promising for miniaturized nanoelectronics, optoelectronic, and flexible device applications [4,6].

Among 2D layered materials, post-transition metal chalcogenides (PTMCs) of the form $TlGaX_2$ (X=S, Se) and $TlInS_2$ have garnered increasing interest due to their distinctive structural, optoelectronic, and electrical properties [9]. These PTMCs exhibit various phase transitions and FE behaviors [10–12], with ferroelectricity proposed to stem from the off-center displacement of $Tl^{1+}$ ions with stereochemically active $6s^2$ lone-pair electrons [12,13]. Among these PTMCs compounds, 2D layered $TlGaS_2$ is of particular interest due to its unique temperature-dependent phase transitions, thermoelectric, and optoelectronic properties [14]. X-ray diffraction (XRD) studies carried out for isomorphic compounds $TlGaS_2$ [15] and $TlGaSe_2$ [16] revealed strictly periodic two-dimensional layers formed by $Ga_4S_{10}$ units consisting of four elementary $GaSe_4$ tetrahedra. The Tl atoms are in trigonal prismatic voids resulting from the combination of the $Ga_4S_{10}$ units into a layer (see Fig. 1(a)). The layer has the symmetry operations corresponding to the $D_{2d}$ point group. Several phase transitions below 300 K have been reported for $TlGaS_2$ [9], however, the information on these phase transitions obtained via different experimental characterization techniques remains inconsistent, and even different authors have derived conflicting results using the same method, as summarized in a review [9]. Notably, despite being isostructural to the FE materials $TlGaSe_2$ and $TlInS_2$ and containing also $Tl^{1+}$ ions with stereochemically active $6s^2$ lone pairs, $TlGaS_2$ has been reported to lack soft modes and FE behaviors [9], which requires further investigation and clarification.

In this work, we investigated in detail the dielectric, heat capacity, and optical properties of $TlGaS_2$ crystals. Intriguingly, we observed a soft-mode behavior, the coexistence of in-plane and out-of-plane polarization, and multiple phase transitions in $TlGaS_2$ crystals.

## II. EXPERIMENTAL DETAILS

The $TlGaS_2$ crystals studied in this work were grown by the Bridgman method in evacuated quartz ampoules, as described in Ref. [17]. Energy dispersive spectroscopy (EDS) measurements confirmed that the chemical composition of the crystals is close to Tl:Ga:S=1:1:2. The dielectric measurements were performed using an LCR meter (TH2828S) with a home-made probe integrated into a Quantum Design (QD) magnetic property measurement system. The RT XRD patterns were recorded using a Philips X'pert PRO X-ray diffractometer with Cu $K_\alpha$ radiation. Specific heat ($C_p$) measurements were carried out in a QD physical properties measurement system. For transmission measurements, thin sample plates were cleaved from the crystals along the (001) cleavage plane using adhesive tape. Thus, all transmission spectra were measured for the polarization direction of the incident light with **E, H** ⊥ $c$. The spectra were recorded in the frequency range 13.5–700 cm$^{-1}$ using a high-resolution Bruker IFS 125 HR spectrometer. The spectral resolution was 0.5 cm$^{-1}$ in the range 13.5–100 cm$^{-1}$ and 0.8 cm$^{-1}$ in the range 100–700 cm$^{-1}$. A liquid-helium-cooled bolometer was used as the detector.

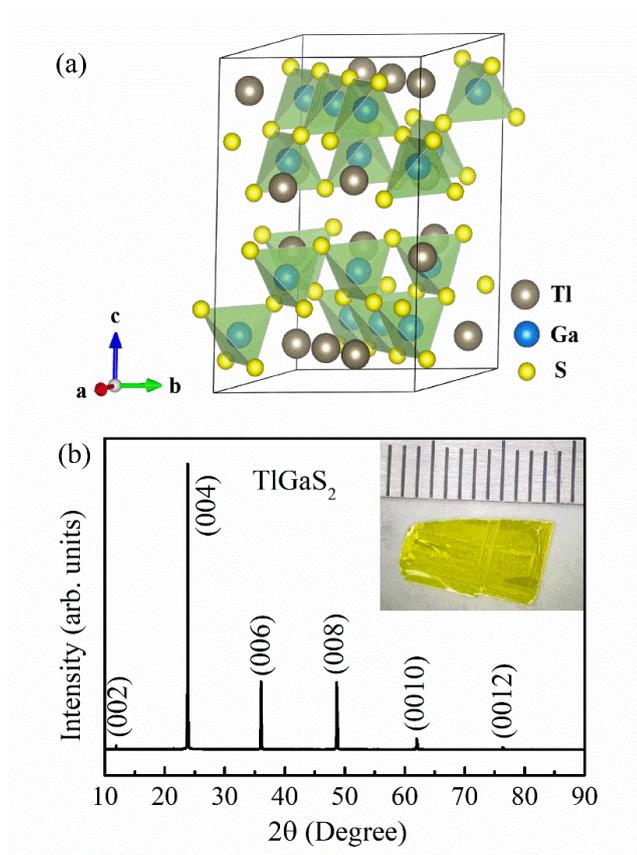

FIG. 1. (a) The crystal structure for the 2D layered TlGaS$_2$. (b) The (00$l$) diffraction patterns for the TlGaS$_2$ single crystal. Inset shows a photograph of the crystal.

### III. RESULTS AND DISCUSSION
#### A. X-ray characterisation

Fig. 1(b) shows the RT XRD pattern of the natural facet of one piece of TlGaS$_2$ single crystal. The sharp and narrow (00$h$) diffraction peaks in Fig. 1(b) suggest the excellent crystallinity of the crystal and *c*-axis orientation of this natural facet. TlGaS$_2$ crystals are yellow in color, as shown in the inset of Fig. 1(b). The facile cleavage along the (001) plane of the crystal is consistent with the 2D layered structure of TlGaS$_2$.

## B. Dielectric measurements

Figures 2 and 3 show the temperature-dependent relative dielectric constant ($\varepsilon'$) along the *c* axis and *ab* plane (i.e., $\varepsilon'_c$ and $\varepsilon'_{ab}$ respectively) for the TlGaS$_2$ crystal in the temperature range of 2– 300 K. $\varepsilon'_c$ decreases initially with decreasing temperature, then increases slightly below $T$~60 K, and gradually becomes nearly constant (~12.8) below $T$~10 K. In constrast, $\varepsilon'_{ab}$ increases continuously with decreasing temperature below 300 K, and tends to saturate at ~70.9 below ~10 K. The significant difference in the $\varepsilon'_{ab}(T)$ and $\varepsilon'_c(T)$ behaviors evidently shows a strong anisotropy in TlGaS$_2$ crystal. Notably, the values of loss tangent (tan δ) for both the *ab* plane and *c* axis are in the range of ~0.001–0.003 below ~200 K, as shown in the right inset of Fig. 2. The low tan δ clearly indicates a negligible contribution from electrical conduction to both in-plane and out-of-plane dielectric constant in the TlGaS$_2$ crystal.

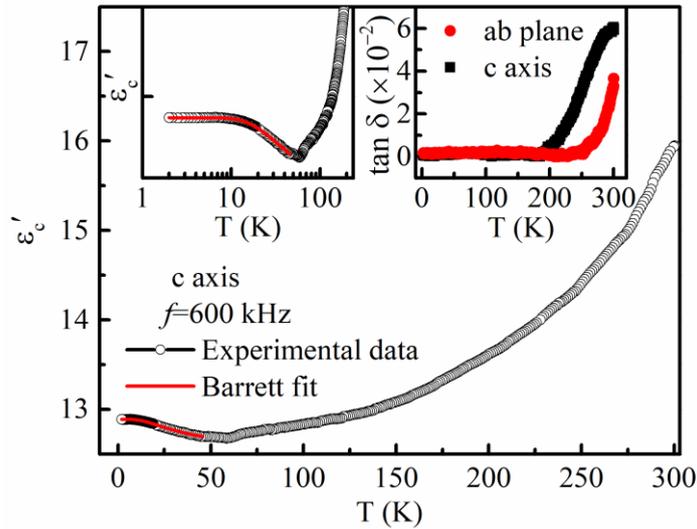

FIG. 2. The dielectric constant $\varepsilon'$ as a function of temperature along the *c* axis for TlGaS$_2$. The red line is a fit to the Barrett relation (1). Left inset is the magnified $\varepsilon'_c(T)$ plot on a logarithmic temperature scale. Right inset shows the loss tangent along the *c* axis and *ab* plane at a frequency of 11 kHz for the crystal.

The increase and subsequent saturation of both $\varepsilon'_c$ and $\varepsilon'_{ab}$ with decreasing temperature at low temperatures in TlGaS$_2$ are important characteristics of typical quantum paraelectrics (i.e., incipient ferroelectrics), such as SrTiO$_3$, KTaO$_3$, and BaFe$_{12}$O$_{19}$ [18–20]. In quantum paraelectrics, the long-range FE order is suppressed to zero temperature by quantum fluctuations associated with zero-point vibration or tunnelling excitations. The saturation behavior of low-T dielectric constant in quantum paraelectrics is in contrast to the sharp or broad dielectric peaks observed near FE or relaxor FE phase transitions in normal FEs or relaxor FEs, respectively.

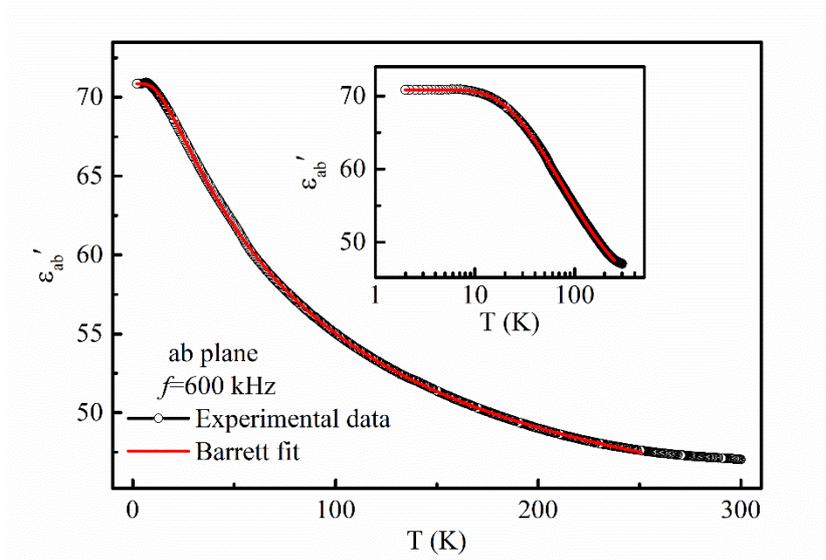

FIG. 3. The dielectric constant $\varepsilon'_{ab}$ as a function of temperature along the *ab* plane for TlGaS$_2$. The red line is a fit to the Barrett relation (1). The inset shows a magnified plot on a logarithmic temperature scale.

The temperature-dependent ε′ behavior (i.e., increase followed by saturation at low *T* on cooling) in quantum paraelectrics obeys well the quantum-mechanical Barrett relation [21],

$$\varepsilon' = A + \frac{C}{\left(\frac{T_1}{2}\right)\coth\left(\frac{T_1}{2T}\right) - T_0}, \quad (1)$$

where *A* and *C* are constants, $T_0$ is the Curie-Weiss temperature in the classical limit, and $T_1$ is the onset temperature of quantum fluctuations. Figs. 2 and 3 together with their insets show the fits of the $\varepsilon'_{ab}(T)$ (2–200 K) and $\varepsilon'_c(T)$ (2-47 K) experimental data to the Barrett relation (1), respectively. The good agreement between the experimental data and the fitted curves for both the $\varepsilon'_{ab}(T)$ and $\varepsilon'_c(T)$ evidently implies a quantum paraelectric behavior and existence of electric dipole interactions along both the *ab* plane and *c* axis in the TlGaS$_2$ crystal. The best-fit parameters are summarized in Table I, along with the parameters for other well-known quantum paraelectrics.

TABLE I. The parameters $C$, $T_0$, and $T_1$ obtained by fitting to relation (1) for TlGaS$_2$. The related parameters for other typical quantum paraelectrics are also listed for comparison.

| Compound | $C$ (K) | $T_1$ (K) | $T_0$ (K) | Refs. |
|---|---|---|---|---|
| SrTiO$_3$ | 90000 | 84 | 38 | [20] |
| EuTiO$_3$ | 23400 | 162 | -25 | [22] |
| NaMnF$_3$ | 3650 | 163 | -7.1 | [23,24] |
| BaFe$_{12}$O$_{19}$ | 452 | 54.9 | -11.7 | [19] |
| DyCrO$_3$ | 2.3 | 208.8 | 89.9 | [25] |
| TlGaS$_2$ (*ab* plane) | 2489 | 41.9 | -58.3 | This work |
| TlGaS$_2$ (*c* axis) | 11.1 | 49.0 | -3.2 | This work |

The negative values of $T_0$ for both the *ab* plane and *c* axis indicate antiferroelectric interactions between the electric dipoles in TlGaS$_2$. The value of $C$ for the *ab* plane of TlGaS$_2$ is comparable to that for NaMnF$_3$, but one order of magnitude smaller than that of SrTiO$_3$. The $T_1$ values in TlGaS$_2$ are smaller than those of SrTiO$_3$ and EuTiO$_3$, but comparable to that of the quantum paraelectric BaFe$_{12}$O$_{19}$ [19,20,26]. Considering that the parameter $C$ in relation (1) can be expressed as $C = n\mu^2/k_B\varepsilon_0$ [21,27], where $n$ is the density of electric dipoles, μ is the dipole moment, and $k_B$ and $\varepsilon_0$ are the Boltzmann constant and permittivity of free space, respectively, the nearly two orders of magnitude larger $C$ along the *ab* plane than along the *c* axis indicates a much larger in-plane electric dipole moment relative to the out-of-plane moment in the TlGaS$_2$ crystal. This larger in-plane dipole moment is also in agreement with the significantly higher $\varepsilon'_{ab}$ than $\varepsilon'_c$ at low temperatures, as shown in Figs. 2 and 3.

The coexistent in-plane and out-of-plane electric dipoles, which was also observed previously in 2D layered semiconductor In$_2$Se$_3$ [28], suggests important potential applications of TlGaS$_2$ in FeFET or FE memory related devices. The in-plane electric dipoles in TlGaS$_2$ are consistent with the theoretical calculation on the origin of ferroelectricity in the isostructural TlGaSe$_2$ [13]. In contrast to the theoretical calculation [13], the presence of additional out-of-plane electric dipoles in TlGaS$_2$ may imply a slightly different lattice structure compared with TlGaSe$_2$, as further confirmed by the infrared (IR) spectroscopy results presented below.

### C. Infrared spectroscopy

In order to gain deeper insight into the possible soft modes and phase transitions, we measured IR spectra of the TlGaS$_2$ crystal at different temperatures. IR spectroscopy is a unique and sensitive technique for identifying local structures and phase transitions in dielectrics.

2D layered TlGaS$_2$ was reported to crystallize in a monoclinic structure with space group $C_{2h}^6$ [15,29]. However, IR reflection and Raman scattering data could not be explained on the basis of this space symmetry group [30]. The authors of Ref. [30] noted, that there are four stacking types of neighboring layers, and in a real crystal a 1D disorder associated with an ambiguity in the layer stacking is produced along the *c* axis perpendicular to the layer plane. By substituting the real disordered crystal by a model ordered crystal with *D*$_{4h}$ symmetry of its primitive cell, a kind of space-based averaging of the possible versions of layer stacking was proposed [30]. In this case, the distribution of vibrational modes according to irreducible representations has the form:

$\Gamma = 7A_{1g} + 3A_{2g} + 8B_{1g} + 2B_{2g} + 14E_g + 2A_{1u} + 8A_{2u} + 3B_{1u} + 7B_{2u} + 14E_u.$

Here, $(A_{2u} + E_u)$ are acoustic modes, $(B_{1g} + E_g)$ are interlayer modes, $7A_{2u}(z) + 13E_u(x,y)$ modes are IR-active. The mode pairs $A_{2u} - B_{2g}$ and $E_u - E_g$ are Davydov doublets [31].

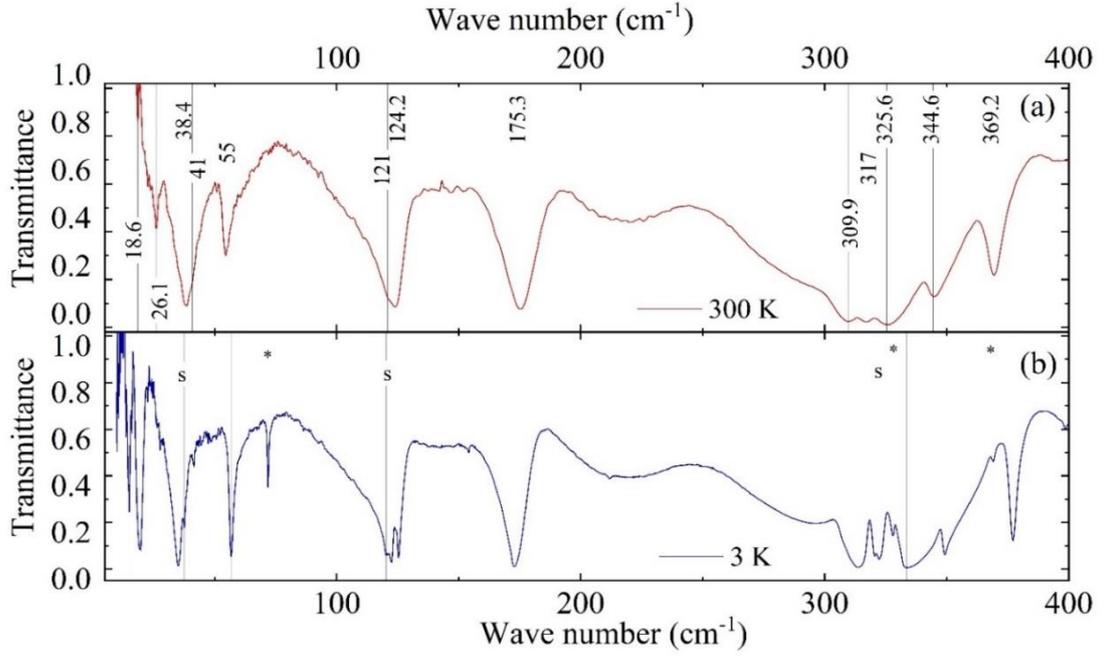

FIG 4. Transmission spectra of TlGaS$_2$ at (a) 300 K and (b) 3 K. Vertical lines indicate the frequencies of all observed phonons. The labels above the vertical lines correspond to the frequencies of 13 phonon modes observed at room temperature. New lines are marked by asterisks. Split lines are labeled with "s".

Figures 4(a) and 4(b) present the IR transmission spectra of the TlGaS$_2$ crystal at temperatures of 300 K and 3 K, respectively. At RT, 13 $E_u$ phonon modes were observed in the TlGaS$_2$ crystal for the light polarization used (k ∥ c, **E, H** ⊥ c – unpolarized light incident perpendicular to the layers' plane), in agreement with the group-theoretical analysis for a nearly tetragonal primitive cell [30] and our previous observation of 13 modes in the isostructural TlGaSe$_2$ crystal [32]. Previous IR reflection studies reported only four or eight phonon modes for TlGaS$_2$ [17,30,33], compared to 13 in this work. The discrepancy between the phonon frequencies reported in Refs. [17,30,33] and in the present study does not exceed 5 cm$^{-1}$. We were able to obtain more detailed information due to a wider spectral range, with a lower frequency limit of 13 cm$^{-1}$ compared to 20 or 60 cm$^{-1}$ in previous works [17,30]. In addition, weak modes are more clearly visible in transmission spectra than in reflection spectra. In this study, weak modes were identified using detailed temperature-dependent spectral analysis. The phonon mode parameters (frequency, linewidth, and intensity) show pronounced temperature dependence, allowing weak lines to be distinguished from the noise. The frequencies of all observed phonon modes at 300, 100, 60, and 3 K are listed in Table II, together with transverse optical (TO) and several Raman mode frequencies reported in Refs. [30] and [33]. Adding the data of Ref. [30] on Raman-active modes, the following Davydov doublets can be distinguished: 24.8 cm$^{-1}$ ($E_\text{u}$) + 22.5 cm$^{-1}$ ($E_\text{g}$); 38.4 cm$^{-1}$ ($E_\text{u}$) + 42.5 cm$^{-1}$ ($E_\text{g}$); 124.2 cm$^{-1}$ ($E_\text{u}$) + 121 cm$^{-1}$ ($E_\text{g}$); 175.3 cm$^{-1}$ ($E_\text{u}$) + 175 cm$^{-1}$ ($E_\text{g}$); 317 cm$^{-1}$ ($E_\text{u}$) + 326 cm$^{-1}$ ($E_\text{g}$); 344.6 cm$^{-1}$ ($E_\text{u}$) + 345.5 cm$^{-1}$ ($E_\text{g}$).

TABLE II. Frequencies (cm$^{-1}$) of phonon modes determined from transmission spectra of TlGaS$_2$ at temperatures of 300, 100, 60, and 3 K in comparison with TO frequencies (cm$^{-1}$) determined from the RT reflectance spectra [30] and [33]. Frequencies of new modes that appear in the spectra at temperatures around 120 K are shown in bold. Subscripts "w" denotes approximate frequencies of weak modes or lines that observed in the spectra as shoulders. Subscripts "s" denotes components of modes split at temperatures of 100–120 K.

|    | This work | | | | Ref. [30] | Ref. [33] |
|----|-----------|------|------|------|-----------|-----------|
|    | 3 K | 60 K | 100 K | 300 K $E_\text{u}$ | 300 K | 300 K $E_\text{u}$ |
| 1  | 15.2 | 16.9 | 18.6 | 18.6 | | |
| 2  | 19.4 | 21.6 | 22.8 | 26.1 | 22.5 $E_\text{g}$ | |
| 3  | 35.1$^\text{s}$ | 35.6$^\text{s}$ | 36.4 | 38.4 | 40 $E_\text{u}$ | |
| 4  | 37.6$^\text{s}$ | 37.5$^\text{s,w}$ | | | 42.5 $E_\text{g}$ | |
| 5  | 41.5 | 41.3$^\text{w}$ | 41.1$^\text{w}$ | 41$^\text{w}$ | | |
| 6  | 56.9 | 56.5 | 56.1 | 55.0 | 55.5 $E_\text{u}$ | |
| 7  | **72.0** | **72.0** | **71.9$^\text{w}$** | - | | |
| 8  | 125.4 | 125.3 | 125.2 | 124.2 | 121 $E_\text{u}$ | 120 |
| 9  | 122.3$^\text{s}$ | 122.5$^\text{s}$ | 122 | 121.0$^\text{w}$ | 121 $E_\text{g}$ | |
| 10 | 120.4$^\text{s}$ | 121.2$^\text{s}$ | | | | |

| 11 | 172.8 | 173.5 | 174.0 | 175.3 | 172 $E_u$ | 174 |
| | | | | | 175 $E_g$ | |
| 12 | 313.3 | 313.3 | 313.1 | 309.9 | 302 $E_u$ | |
| 13 | 320.6$^s$ | 320.1$^{s,w}$ | | | | |
| 14 | 322.4$^s$ | 321.3$^s$ | 320.7 | 317.0 | 320 $E_u$ | 317 |
| | | | | | 326 $E_g$ | |
| 15 | **327.9** | **327.2$^w$** | **327.0$^w$** | - | - | |
| 16 | 333.7 | 332.9 | 332.2 | 325.6 | - | |
| 17 | 349.2 | 349.0 | 348.6 | 344.6 | 341 $E_u$ | |
| | | | | | 345.5 $E_g$ | |
| 18 | **368.9** | **368.5$^w$** | **367.8$^w$** | - | 360 $A_{2u}$ | |
| 19 | 377.1 | 376.1 | 375.1 | 369.2 | 364 $E_u$ | 367 |

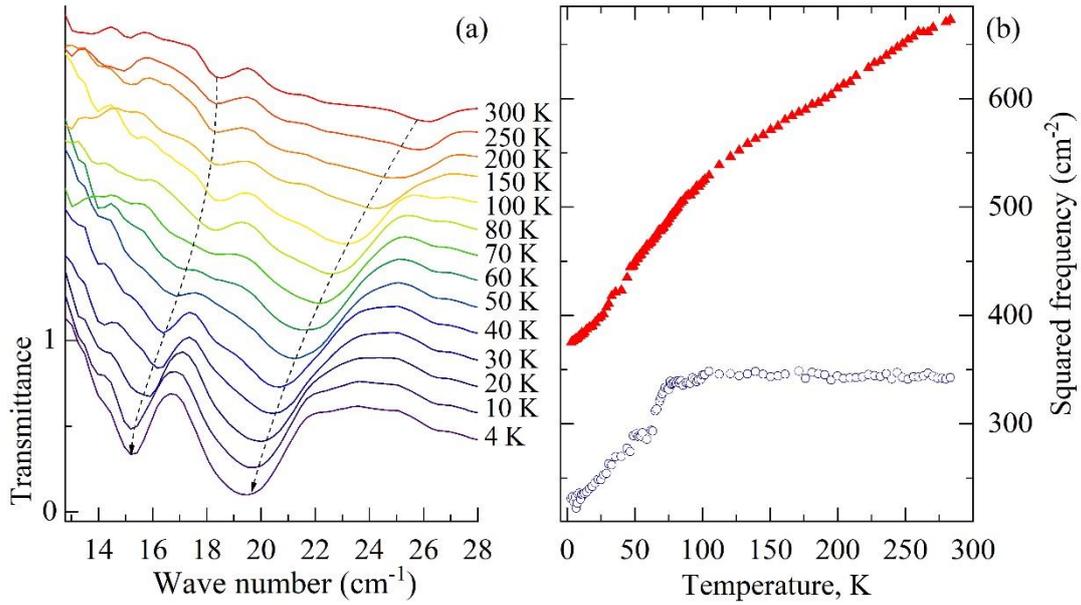

FIG. 5. (a) Transmission spectra of TlGaS$_2$ crystal in the spectral range 13.5–28 cm$^{-1}$ at different temperatures. Here and throughout, the transmission spectra are vertically offset with respect to each other. The transmission scale is shown for the spectrum at 3 K; the same scaling is preserved for all other spectra. The dashed arrows are guides to the eye. (b) Temperature dependence of the squared frequencies of the two modes shown in Fig. 5(a).

The IR spectra of TlGaS$_2$ exhibit intriguing temperature-dependent behavior, which is discussed in detail below for different frequency ranges. Figure 5(a) shows the IR spectra in the range 13–28 cm$^{-1}$ at different temperatures. Two phonon modes with frequencies of 18.6 and 26.1 cm$^{-1}$ were observed at RT. The temperature dependences of the squared frequencies of the lines are shown in Fig. 5(b). Upon cooling, the 26.1 cm$^{-1}$ mode narrows and monotonically shifts to lower frequencies, with a much weaker frequency reduction below ~20 K. In contrast, the phonon mode near 18.6 cm$^{-1}$ remains nearly unchanged down to $T$ ~75 K, then it begins to soften sharply, and finally tends to level off gradually below ~15 K, as shown Figs. 5(b) and 6. The slightly increased noise in the phonon frequency below ~12 cm$^{-1}$ arises from measurements near the frequency limit of our spectrometer. The anomaly near $T$ ~75 K for the 18.6 cm$^{-1}$ polar soft mode and the dielectric minimum in $\varepsilon'_c(T)$ near $T$~60 K, both of which are associated with the vibration of Tl$^{1+}$ ions, may indicate a possible local lattice distortion related to the emergence of out-of-plane electric dipoles in TlGaS$_2$.

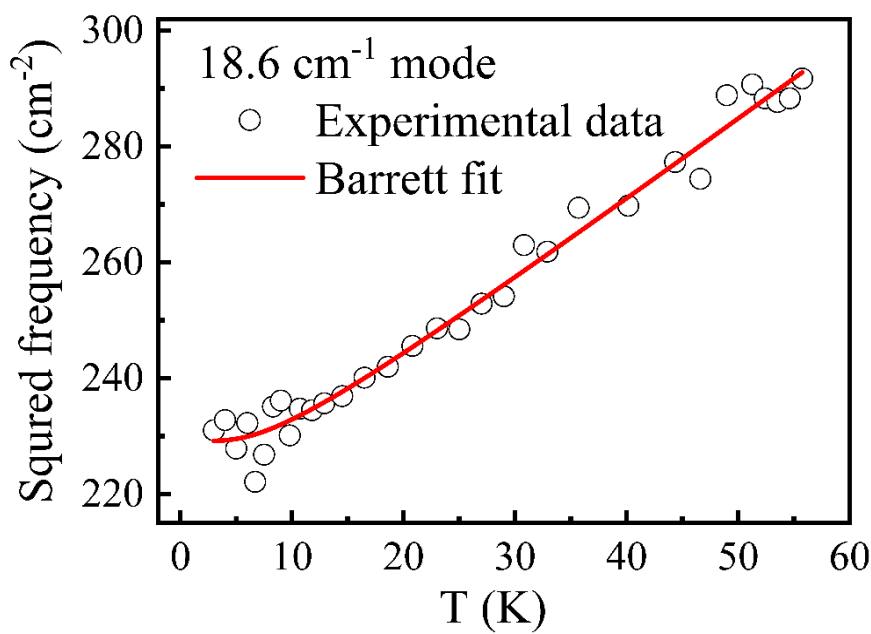

FIG. 6. The squared frequency of the soft phonon mode near 18.6 cm$^{-1}$ as a function of temperature. The red solid line is a fit to the squared Barrett formula.

This freezing-like soft-mode behavior is another hallmark of quantum paraelectrics [26,34]. In the case of quantum paraelectrics such as SrTiO$_3$, quantum fluctuations suppress the FE order and stabilize the paraelectric state even at zero temperature, so that the soft-mode frequency does not reach zero [34]. In contrast, the frequency of soft modes in classical FEs decreases linearly and approaches zero as the FE phase transition temperature is approached [34,35]. We fitted temperature-dependent squared frequency of the polar soft mode near 18.6 cm$^{-1}$ using the squared Barrett formula [26,34], $\omega^2 = C'[(\frac{T_1}{2})\coth(\frac{T_1}{2T}) - T_0]$, where ω is the frequency of the phonon mode. As can be seen in Fig. 6, the calculated curve is in relatively good agreement with the experimental data, yelding best-fit parameters $C'$~1.4 cm$^{-2}$ K$^{-1}$, $T_0$ ~ −151.6 K, and $T_1$ ~ 22.6 K. The signs of $T_1$ and $T_0$ are consistent with those obtained from the dielectric constant data (Fig. 2), but their absolute values differ by a factor of 2–3.

The line intensity of the 18.6 cm$^{−1}$ mode shows non-monotonic temperature dependence. Near $T$ ~75 K, the 18.6 cm$^{−1}$ mode broadens and becomes almost indistinguishable, while its intensity increases again upon further cooling. A FE soft mode was observed previously in the isostructural TlGaSe$_2$ crystal by using submillimeter dielectric spectroscopy [10,11]. The soft-mode frequency approached zero near the FE phase transition at $T_c$ = 107 K. The mode around 18 cm$^{-1}$ at $T$ = 80 K was also detected in the submillimeter spectra of TlGaS$_2$ [36], however, its behavior upon further cooling was not studied.

The phonon modes at low frequencies, including the soft modes near 18.6 and 26.1 cm$^{−1}$, in TlGaS$_2$ can be ascribed to the vibration of the heavy Tl$^{1+}$ ions and/or translational motions of the Ga$_4$S$_{10}$ polyhedra [30]. This Tl$^{1+}$ ions related soft-mode behavior in quantum paraelectric TlGaS$_2$ is consistent with the off-center displacement of Tl$^{1+}$ ions induced electric polarization in the isostructural FE TlGaSe$_2$ and TlInS$_2$ [13].

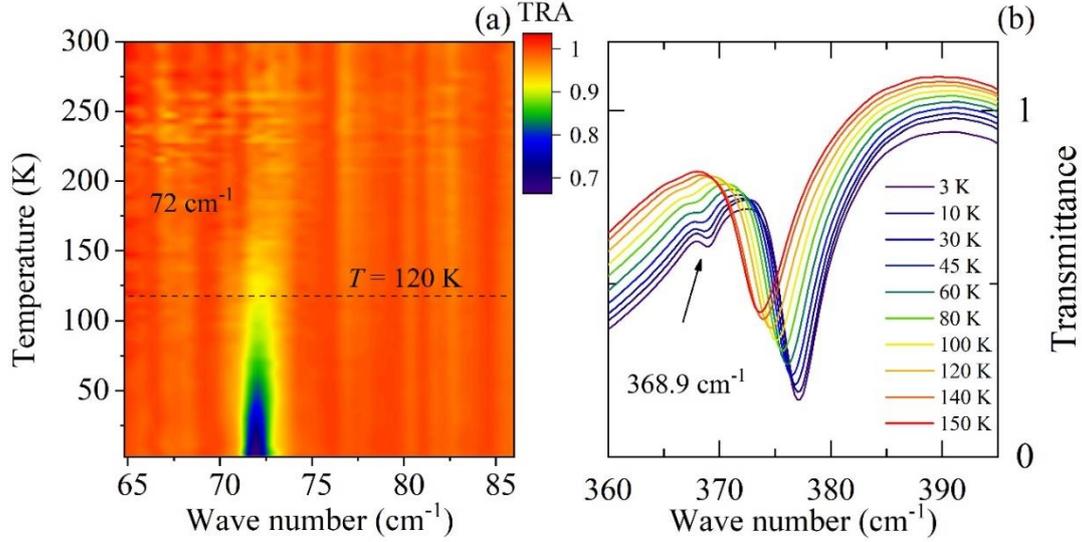

FIG. 7. Transmission spectra of TlGaS$_2$ crystal (a) presented as an intensity map in the temperature – frequency (60–85 cm$^{-1}$) axes; (b) at different temperatures in the frequency range 360–395 cm$^{-1}$.

In addition to the soft-mode behavior of the two low-frequency modes, we observed other peculiarities in the temperature-dependent IR spectra of TlGaS$_2$ crystal, namely, the appearance of new lines and a distinct splitting of some phonon modes, which unambiguously points to structural changes. At a temperature of about 120 K, three weak new lines appear, the intensity of which increases with further cooling. At $T$ = 3 K, their frequencies are 72 cm$^{-1}$ (see Fig. 7(a)), 327.9 cm$^{-1}$ (see Fig. 8(a)), and 368.9 cm$^{-1}$ (see Fig. 7(b)), respectively.

The limited amount of information available in the literature on phase transitions in TlGaS$_2$ remains rather contradictory [9]. During low-temperature XRD studies of a single crystal [37], anomalies (jumps) in the temperature dependences of the lattice parameters *a*, *c*, and *β* were detected at $T$ = 121±1 K, indicating a possible structural phase transition. The authors noted that the lattice distortion at the phase transition is apparently so small that this transition had not previously been detected in polycrystalline samples. A transition near 120 K was also suggested based on optical absorption data [38]. In addition, a number of anomalies were found in heat-capacity measurements at $T$ = 73.5 K and at higher temperatures [39]. It should be noted that this finding contradicts the results of Ref. [40], where no anomalies were found in the temperature dependence of the heat capacity of TlGaS$_2$. Raman spectroscopy data revealed the appearance of new modes at a temperature of 180 K [14,41] and a splitting of some modes at 250 K [41]. However, comparison of the spectra presented in Ref. [41] at temperatures of 130 K, 75 K, and at the lowest investigated temperature of 8.5 K (Fig.3 in [41]) shows that new lines apparently appear at temperatures below 130 K.

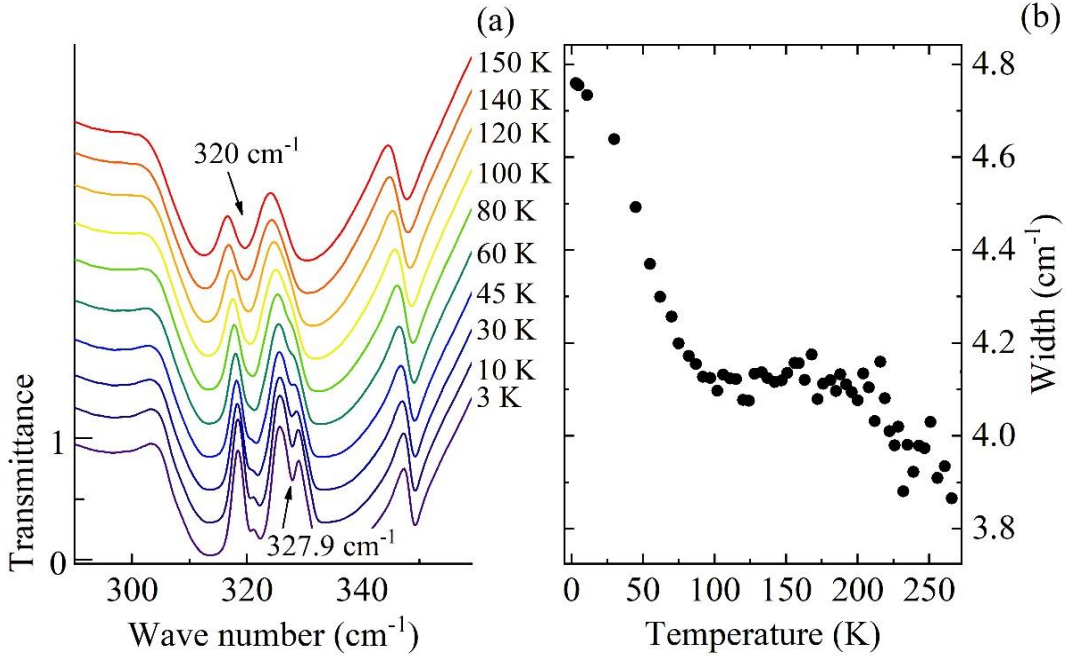

FIG. 8. (a) Transmission spectra in the frequency range 290–360 cm$^{-1}$ at different temperatures; (b) temperature dependence of the linewidth of the 320 cm$^{-1}$ mode.

It should be noted that in the isostructural TlGaSe$_2$ crystal, together with FE ordering, a phase transition to an incommensurately modulated structure occurs at $T_i$ = 120 K, preceding a transition at $T_c$ = 107 K to a commensurate structure with a quadrupled unit cell. The transition at $T_i$ is accompanied by the activation of a large number of lines in the IR and Raman spectra of TlGaSe$_2$ (for example, we observed the appearance of 7 new modes in the IR spectra for the $E \perp c$ polarization direction of the incident light [32]). The spectral changes near $T \sim 120$ K in TlGaS$_2$ are much less pronounced than in TlGaSe$_2$ crystal, which may support the assumption [37] of only minor structural changes at the phase transition.

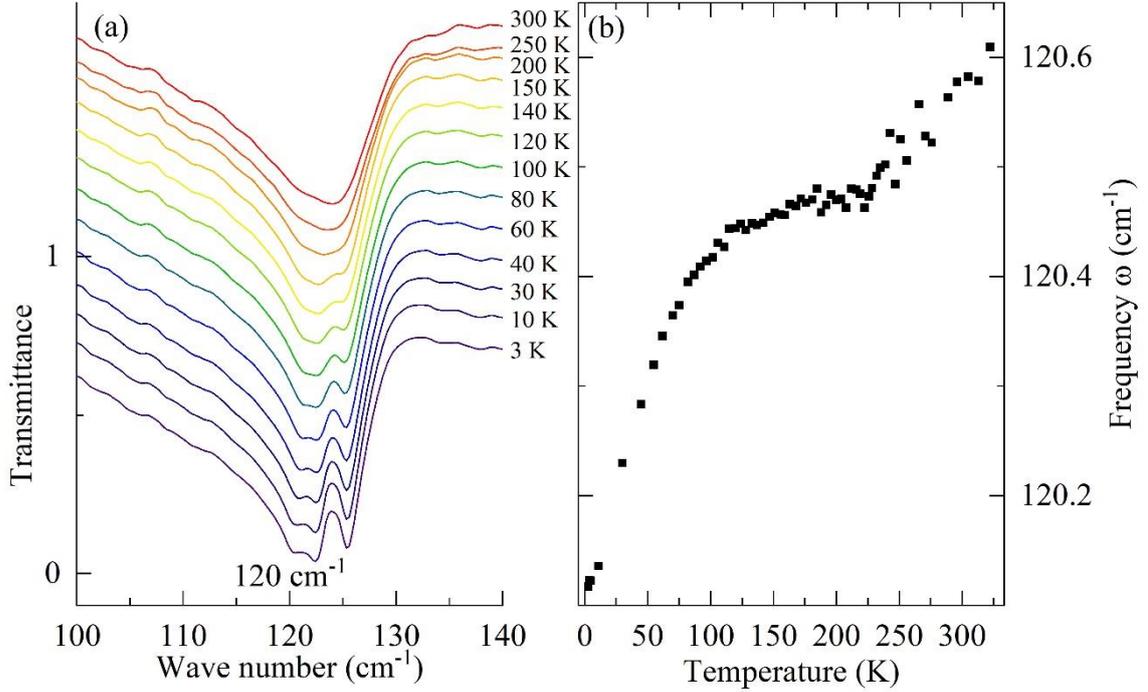

FIG. 9. (a) Transmission spectra in the frequency range 100–140 cm$^{-1}$ at different temperatures and (b) temperature dependence of the frequency of the 120 cm$^{-1}$ mode.

In addition to the appearance of new lines upon cooling, the spectra also show a splitting of three modes, which at RT have frequencies of 38.4, 121, and 317 cm$^{-1}$. Figures 8(a) and 9(a) show IR spectra measured at different temperatures that illustrate the splitting of the 317 and 121 cm$^{-1}$ modes, respectively. The components of the split modes are separated by a small frequency gap comparable to their widths, which makes it difficult to determine the exact temperature at which the splitting occurs. The analysis of temperature dependences of the mode parameters is further complicated by the high concentration of modes in the spectral ranges under discussion. For example, in the frequency range 300–360 cm$^{-1}$, there are four phonon modes at RT and six modes at $T = 3$ K. The spectral line near 120 cm$^{-1}$ at RT has a complicated shape. Upon decreasing temperature, a separation of the low-frequency component is observed, which transforms into a distinguishable line and then undergoes splitting. The split components can be clearly resolved only at the lowest temperatures. However, by considering the split component as a whole, additional information can be extracted from the temperature dependence of its parameters. Figure 9(b) shows the temperature dependence of the gravity center of the mode 120 cm$^{-1}$ having a complicated structure. A pronounced low-frequency shift is observed below 100–120 K. The 317 cm$^{-1}$ line begins to broaden rapidly just in this temperature range (see Fig. 8(b)), indicating the onset of splitting.

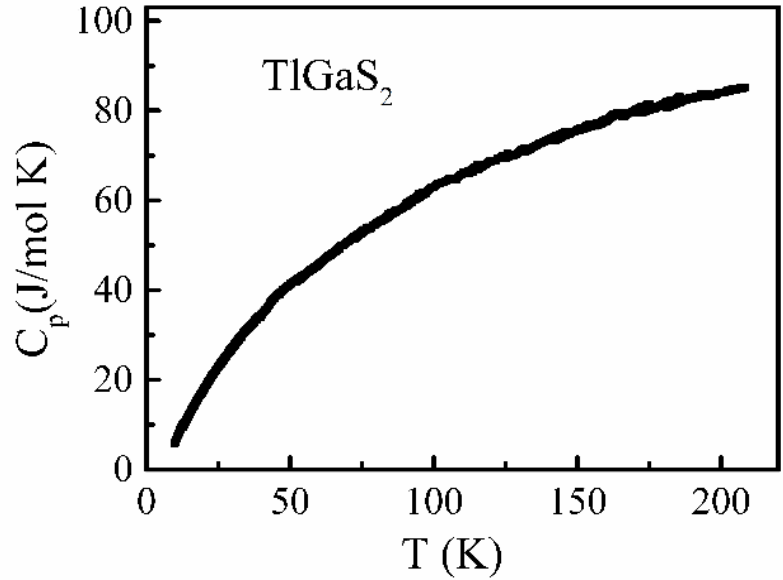

Fig. 10 The temperature-dependent heat capacity for the TlGaS$_2$ crystal.

The splitting of phonon modes indicates changes in the crystal structure. In the present case, this splitting may be associated with both the phase transition at 120 K and further structural transformations at lower temperatures. In order to further investigate possible phase transitions in the TlGaS$_2$ crystal, we measured the temperature dependence of the heat capacity $C_p$ ($T$) in the temperature range 10–200 K. As shown in Fig. 10, $C_p$ does not show any distinguishable anomalies in the this temperature range, including near $T$~75 and ~120 K, agreement with the results of Ref. [40]. Importantly, the splitting of the IR phonon lines or emergence of new IR peaks shown in Figs. 7-9 occur without visible anomalies in $C_p$ ($T$) (see Fig. 10), suggesting that the phase transitions indicated by the IR spectra are weak or short-range, rather than strong and long-range. It should be noted that due to the weakness of the interlayer interaction and the ease of formation of various modifications only due to the displacement of layers relative to each other, TlGaSe$_2$ crystals are polytypic and contain defects, which influence phase transitions. This may be the reason for contradictory results in the literature on phase transitions in these crystals.

## IV. CONCLUSIONS

In summary, we have systematically investigated the dielectric, heat capacity, and IR optical properties of 2D van der Waals layered TlGaS$_2$ single crystals. Our results demonstrate the coexistence of larger in-plane and smaller out-of-plane electric dipoles associated with possible off-center displacement of Tl$^{1+}$ ions in TlGaS$_2$. Quantum paraelectric behavior in TlGaS$_2$ was confirmed by the temperature dependence of the dielectric constant, which follows the Barrett relation, and by the freezing-like soft-mode behavior observed in IR spectra. IR spectroscopy reveals the emergence of three new phonon modes near $T \sim 120$ K and splitting of three additional modes, consistent with weak or short-range structural transitions supported by the absence of heat capacity anomalies in the same temperature range. These transitions are less pronounced than those in the isostructural TlGaSe$_2$, suggesting minor structural distortions in TlGaS$_2$. Our findings resolved previous inconsistencies regarding phase transitions and soft modes in TlGaS$_2$ and provide key experimental insights into the coexistence of in-plane and out-of-plane electric dipoles in 2D layered post-transition metal chalcogenides, laying the groundwork for their potential applications in FeFETs and other ferroelectric-based nanoelectronic devices.

## ACKNOWLEDGMENTS

The optical part of the work was carried out within the framework of state assignments of the Ministry of Science and Higher Education of the Russian Federation for the Institute of Spectroscopy Nos. FFUU-2025-0004 (M.N.P.) and FFUU-2024-0004 (A.D.M.). K.R.A. is grateful to the Rector of the National Aviation Academy, Academician A.M. Pashayev, for supporting his work on crystal growth.